\documentclass{PoS}

\title{CNO production in the first generation stars}

\ShortTitle{CNO production in the first generation stars}

\author{\speaker{Sylvia Ekstr\"om}, Georges Meynet \& Andr\'e Maeder\\
        Geneva Observatory, Switzerland\\
        E-mails: \email{sylvia.ekstrom@obs.unige.ch},\\
                      \email{georges.meynet@obs.unige.ch},\\
                       \email{andre.maeder@obs.unige.ch}}

\abstract{Big Bang nucleosynthesis produces only light elements and the very first generation stars are thus formed from metal-free clouds. They start the production of heavy elements during their life, and enrich the interstellar medium through their explosive death. Stellar evolution models show that the treatment of rotation has important effects on the evolution of those metal-free stars: for example, rotating models produce up to five orders of magnitude more primary nitrogen than non rotating models, due to internal mixing. This will have an impact in the composition of the second generation stars, some of which may now be observed in the Galactic halo. In the case Population III stars were very massive and would end up as direct black holes, rotation again have an interesting effect of enhancing mass loss through centrifugal force and surface enrichment. CNO composition patterns observed in ultra metal-poor halo stars may be explained by a 'wind only' contribution.}

\FullConference{International Symposium on Nuclear Astrophysics --- Nuclei in the Cosmos --- IX\\
		 June 25-30 2006\\
		 CERN, Geneva, Switzerland}

\begin{document}
\newcommand{\msol}{\ensuremath{\rm M_{\odot}}}
\newcommand{\vcrit}{\ensuremath{\upsilon_{\rm crit}}}
\newcommand{\lsol}{\ensuremath{\rm L_{\odot}}}

\section{Primordial stars}
\subsection{Birth}
After the Big Bang, the early Universe is composed only from H and He, a little Be, B and Li, but no heavier elements. At the end of the so-called 'Dark Age', halos of dark matter start to collapse and form stars with the gas trapped within. The star-forming clouds are almost pure H and He, so fragmentation cannot occur, leading to a massive or very massive star formation (see Omukai \& Palla 2003 and references therein). The mass domain covered by primordial stars is still very much debated. Numerical simulations including known physics lead to the formation of very massive objects (see Bromm \& Larson 2004), with masses up to 500-700 \msol. Stability studies for such extreme stars show that they are stable enough against radial pulsations to evolve until advanced stages (see Baraffe \& al. 2001). Observational constraints, though, favor a more standard, slightly top-heavy Larson IMF in the range 1 - 100 \msol\ (see Schneider \& al. 2006), while the mechanism for producing sufficient fragmentation in primordial clouds is not yet known.

\subsection{Evolution}
The new star is born metal-free. In its contracting core, the gas heats up until nuclear reactions start burning some H, but the only energy source is provided by pp-chains reactions, so the collapse is only slowed down. The core reaches then a temperature that allows some He burning through 3$\alpha$ reaction, producing some C and allowing the CNO-cycle energy production to stop the collapse. Primordial stars are thus born hotter and more compact, and their metal-free envelope lets them  remaining so through the whole main sequence (MS). Since radiative winds scale with the metallicity as $\dot{M} \sim Z^{0.65}$ (see Vink \& al. 2001), the star experiences practically no mass loss during its evolution, except if its luminosity is high enough to remove some mass just with H or He lines (see Kudritzki 2002).

\subsection{Death}
According to Heger \& al. (2003), the fate of a massive star depends on the mass of its He core at the end of the evolution. We have just seen that primordial stars evolve at almost constant mass, so they end up their life with a heavy He core. Stars with initial mass ranging from 25 to 140 \msol\ are supposed to collapse as direct black hole without ejecting any matter. Between 140 and 260 \msol\ the star undergoes pair-creation instability and is supposed to be completely disrupted in the process. Heavier stars again form direct black hole.


\section{Rotation at (extremely) low metallicity}
A detailed description of the effects of rotation at low or very low metallicity is given in Meynet \& al. (2006). We will highlight here two main effects: 1) because of the very low radiative winds, the star cannot lose its angular momentum and is thus prone to reach the break-up velocity during the MS; 2) because of a less efficient meridional circulation, the $\Omega$-profiles are steeper, and thus rotational mixing is stronger (see also the contribution of G. Meynet in this conference, \pos{PoS(NIC-IX)015}).

The first effect leads to enhanced mass loss through purely mechanical process, the uppermost layers of the star being removed by the centrifugal force. The second effect takes place during core He-burning, enriching the surface with newly synthesized elements and allowing radiative winds to get stronger.

\section{Modelization}
In order to study the effects of rotation at $Z=0$, we have computed, with the Geneva stellar evolution code, metal-free models ranging from 9 to 200 \msol\ (Fig. \ref{fig2}). For each mass, we have computed a non-rotating model and a differentially rotating one, where the equatorial velocity has been chosen so that $\upsilon/\vcrit \simeq 0.5$ (depending on the mass, this gives $\upsilon_{\rm eq} \simeq 500 - 800\ {\rm km\ s^{-1}}$), corresponding to the average value observed in massive stars. The initial composition of the models is $X=0.76$, $Y=0.24$\ and $Z=0$.
\begin{figure}[h]
\begin{center}
\includegraphics[width=.6\textwidth,height=.5\textwidth]{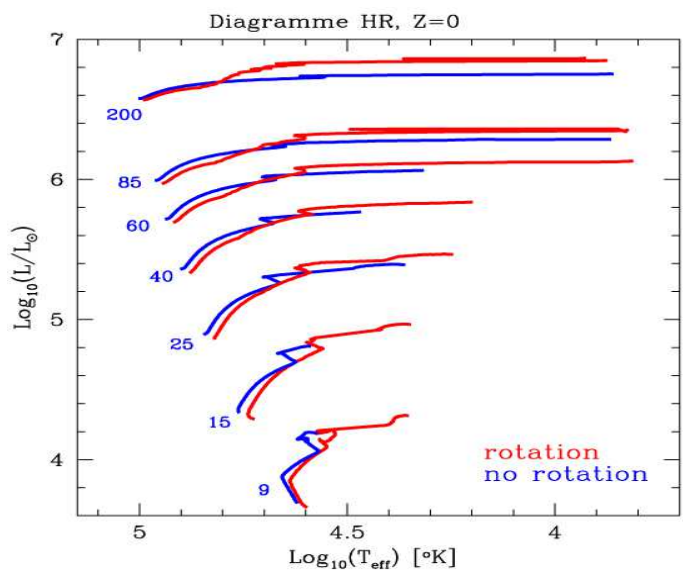}
\caption{Herztsprung-Russell diagram of the $Z=0$\ models}
\label{fig1}
\end{center}
\end{figure} 

The input physics is the same as in Meynet \& Maeder (2002), including rotation-induced instabilities as meridional circulation, secular and dynamical shear, and horizontal turbulence. We have used the Schwarzschild criterion for convection, and an overshooting parameter of $0.2 {\rm H_{P}}$\ during core H- and He-burning and 0 otherwise. We have applied the mass loss rates proposed by Kudritzki (2002) for the models reaching a luminosity $L/\lsol \geq 6$, with the same adaptation for the strictly $Z=0$ case as in Marigo \& al. (2003).

\section{CNO production}

The models have been followed until the end of core O-burning, so the values presented here are not the final yields of the model, but they give insights of the way differential rotation modifies the characteristics of a star. As expected, most of the models have reached the break-up velocity during the MS, some of them very early in their evolution. The mass lost within this process however is not very important because only the uppermost layers of the star are removed, and those layers have a low density. Rotation favors a redward evolution and, as seen above, an enhancement of the effective surface metallicity through mixing, so the radiative winds are stronger during He-burning. Again, in the $Z=0$ case, this doesn't lead to strong mass loss because the time involved is short. Our heavier rotating model (200 \msol) loses just a little more than 10\% of its initial mass (8.2\% during MS, 2.2\% after), but its non-rotating counterpart loses all added only 0.5\%.

\begin{figure}[h]
\begin{center}
\includegraphics[width=\textwidth,height=.45\textwidth]{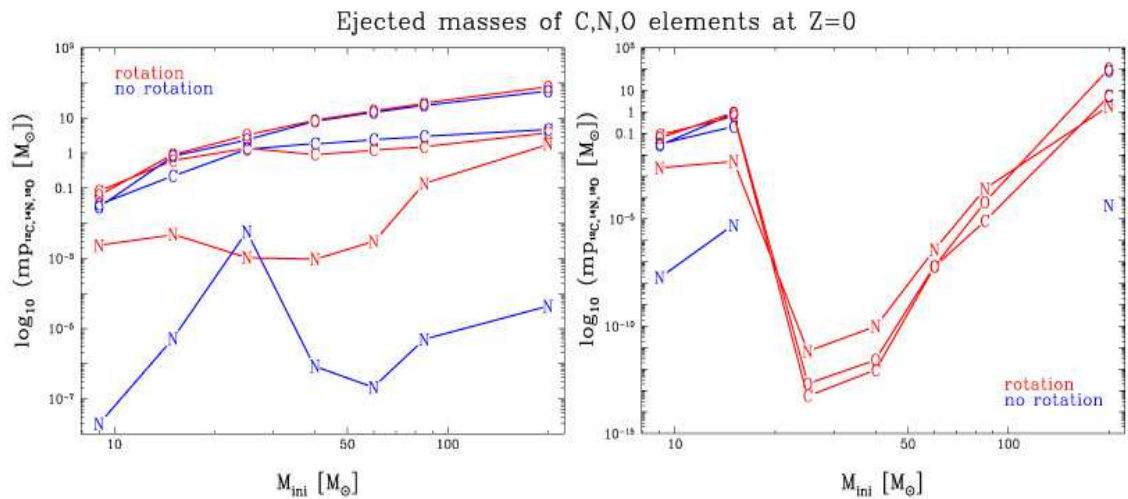}
\caption{CNO production of $Z=0$\ models. \textit{{\bf Left:}} assuming a SN explosion with a remnant mass determined by the CO core mass as in Maeder (1992); \textit{{\bf right:}} following the supposed fate of the star according to Heger \& al. (2003).}
\label{fig2}
\end{center}
\end{figure} 
At the end of the evolution, the CO cores of the rotating models are about 10-20\% heavier than the non-rotating ones. Since the bulk of the CO produced comes from the core, the values obtained for both the rotating and non-rotating models are quite similar (see Fig. \ref{fig2}, left).

The case of N is very different. Except for the 25 \msol\ model, the rotating values are three to five orders of magnitude higher than the non-rotating ones. This is due to the fact that N is mainly produced in the H-burning shell, when some C and O from the He-burning shell is diffused outwards. Rotation-induced mixing favors this process. But for the 25 \msol\ without rotation, as Limongi \& Chieffi (2005), we find a high primary N production about three orders of magnitude higher than the value produced by the other non-rotating models. The value is even slightly larger than the rotating model's value. It can be explained by structure differences: in the non-rotating case, the convective zone associated with the He shell is more extended than in the rotating case, and reaches the H shell, leading to a strong production of primary N.

The picture becomes very different if we take into account the supposed fate of the star according to Heger \& al. (2003). In the non-rotating case, the models with masses between 25 and 85 \msol\ do not contribute at all to the enrichment of the medium, having no winds and ending as direct black holes. In the rotating case, the enrichment is made through a 'wind-only' contribution, leading to low but non-zero contribution of the enrichment of the medium. An important thing to note is that the mass lost through winds has very peculiar abundances compared to the composition of SN-ejecta (see Fig. \ref{fig2}, right), with larger N/O, N/C ratios, and much lower $^{12}$C/$^{13}$C (not shown).

To widen the discussion, we have also computed fast rotating 60 \msol\ models at $Z=10^{-5}$\ and $10^{-8}$\ (see Meynet \& al. 2006). In those models, the effects described above are very strong, leading to an increase of the effective surface metallicity of 2 to 6 orders of magnitude, and a mass loss of 40-60\% of the initial mass. If we compare the wind-only ejecta and the abundances pattern observed in some of the most metal-poor stars of the halo of our Galaxy (see Frebel \& al. 2006 for HE 1327-2326 and Plez \& Cohen 2005 for G77-61), we get an interesting matching able to reproduce the high C, N, and O at the same time (see Fig. 8 of Meynet \& al. 2006).

\end{document}